\documentclass{ws-ijbc}
\usepackage{graphicx}
\usepackage{graphics}
\usepackage{subfigure}
\usepackage{MnSymbol}

\DeclareTextFontCommand{\textswab}{\swabfamily}
\usepackage{xcolor}
\begin{document}

\catchline{}{}{}{}{} 

\markboth{Ala\c{c}am and Shilnikov}{CPGs out of latent parabolic bursters}

\title{MAKING A SWIM CENTER PATTERN GENERATOR OUT OF LATENT PARABOLIC BURSTERS}

\author{DEN\.{I}Z ALA\c{C}AM}
\address{Department of  Mathematics and Statistics, Georgia State University, Atlanta 30303, USA.\\
Email: \texttt{dalacam1@student.gsu.edu.}}

\author{ANDREY SHILNIKOV}
\address{Neuroscience Institute and Department of  Mathematics
and Statistics, Georgia State University, Atlanta 30303, USA. Email: \texttt{ashilnikov@gsu.edu.}\\
Institute for Information Technology, Mathematics and Mechanics, Lobachevsky State University of Nizhni Novgorod, Nizhni Novgorod, 603950,  Russia.  }

\maketitle

\begin{history}
\received{(to be inserted by publisher)}
\end{history}

\begin{abstract}

We study the rhythmogenesis of oscillatory patterns emerging in network motifs composed of inhibitory coupled tonic spiking neurons represented by the Plant model of R15 nerve cells. Such motifs are argued to be used as building blocks for a larger central pattern generator network controlling swim locomotion of sea slug {\it Melibe leonina}.

\end{abstract}



\section{Introduction}\label{sec:1}

A plethora of vital rhythmic motor behaviors, such as heartbeat, respiratory functions and locomotion are produced and governed by neural networks called central pattern generators (CPGs) \cite{CPG,Bal1988,Marder-Calabrese-96,Frost09011996,KCF05,KatzHooper07}. A CPG is a microcircuit of interneurons whose mutually synergetic interactions autonomously generate an array of multi-phase bursting rhythms underlying motor behaviors. There is a growing consensus in the community of neurophysiologists and computational researchers that some basic structural and functional elements must be shared by CPGs in invertebrate and vertebrate animals. As such, we should first understand these elements, find the universal principles, and develop efficient mathematical and computational tools for plausible and phenomenological models of CPG networks. Pairing experimental studies and modeling studies has proven to be key to unlocking insights into operational and dynamical principles of CPGs \cite{Gillner-Wallen-85,Kopell26102004,Ma87,Kopell88,Canavier1994,SKM94,Dror1999,Prinz2003}. Although various circuits and models of specific CPGs have been developed, it still remains unclear what makes the  CPG dynamics so robust and flexible   \cite{Best2005,prl08,SHWG10,Koch01122011,Calabrese01122011,M12}. It is also unclear what mechanisms a multi-functional motor system can use to generate polyrhythmic outcomes to govern several behaviors \cite{Kristan2008b,Briggman2008,Wojcik2014}. Our goal is to gain insight into the fundamental and universal rules governing pattern formation in complex networks of neurons. To achieve this goal, we should identify the rules underlying the emergence of cooperative rhythms in simple CPG networks.  

\begin{figure*}[t!]
\centering
\subfigure [][] {\resizebox*{.56\columnwidth}{!}{\includegraphics{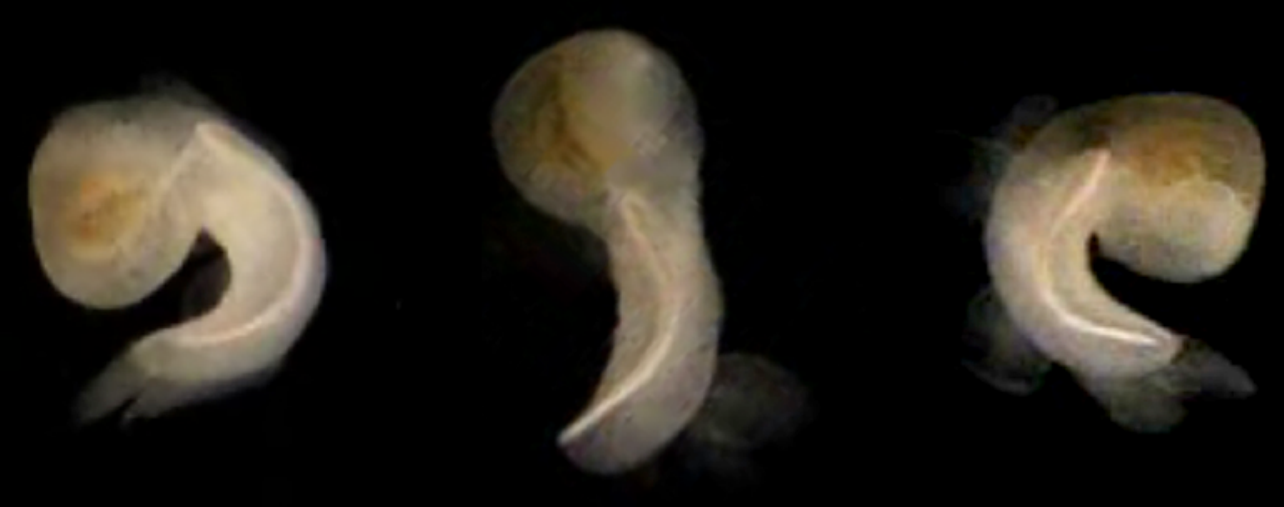}}} ~~~
\subfigure [][] {\resizebox*{.35\columnwidth}{!}{\includegraphics{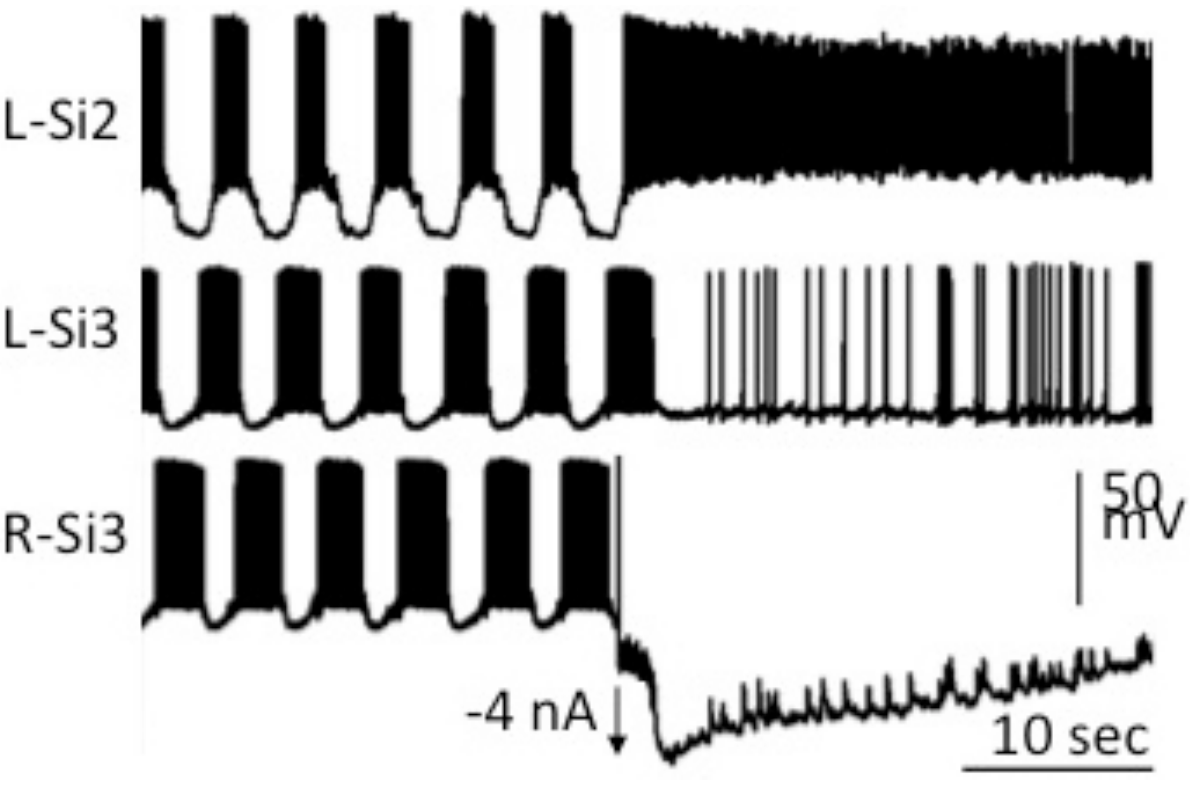}}} 
\caption{(a) {\it Melibe leonina} lateral swim style.
(b) Network bursting in swim interneurons (Si) of the {\it Melibe} swim CPG halts when Si3R is hyperpolarized, thus its counterpart Si3L begins tonic spiking; the photographs and in-vitro recording provided courtesy of A.~Sakurai \cite{sck2014}} \label{fig0a}
\end{figure*}

\begin{figure*}[ht!]
\centering
\subfigure [][] {\resizebox*{.26\columnwidth}{!}{\includegraphics{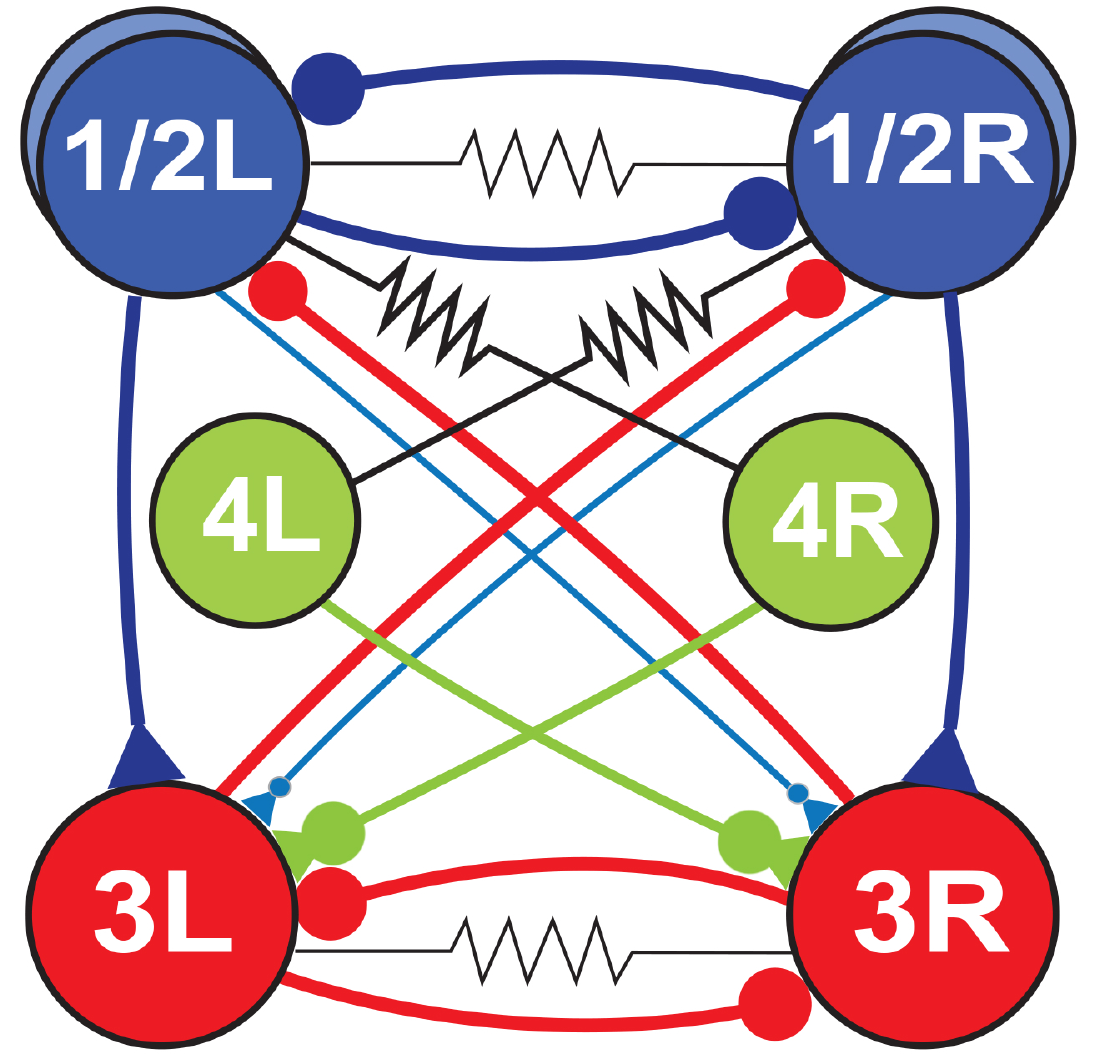}}} ~~~~
\subfigure [][] {\resizebox*{.4\columnwidth}{!}{\includegraphics{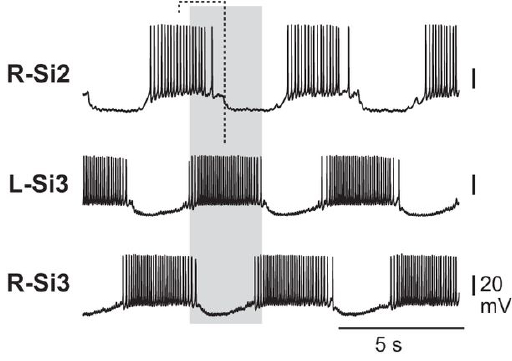}}} 
\caption{(a)   A core circuitry of the biological {\em Melibe} swim CPG  with inhibitory ($\bullet$), excitatory ($\filledmedtriangledown$) and   electrical ($ \backslash\slash\backslash\slash$) synapses  \cite{sck2014}. (b) {\em In-vitro}  voltage activity recordings from identified swim interneurons, Si2L and Si3L/R, of the {\em Melibe} swim CPG with the characteristic $\frac{3}{4}$-phase lag between the HCO2 and HCO3; intracellular recording provided courtesy of A.~Sakurai \cite{sck2014}.}\label{fig0b}
\end{figure*}

Recently, a great deal of computational studies have been focused on a range of 3-cell motifs of bursting neurons coupled by chemical (inhibitory and excitatory) and electrical synapses to disclose the role of coupling in generating  sets of coexisting rhythmic outcomes, see \cite{Shilnikov2008b,Wojcik2011a,Wojcik2014,Schwabedal-Neiman-Shilnikov-14, drake_jus2015,CKAXKS15} and references therein. These network structures reflect the known physiological details of various CPG networks in real animals. Next, we would like to explore dynamics and stability of some identified CPG circuits constituted by 4-cells \cite{jalil2013}. Examples of such sub-networks can be found in the cerebral crustacean stomatogastric ganglion (STG) \cite{CPG,prinz2003functional,prinz2004similar,M12}, as well as in the swim CPGs of the sea slugs --  {\em Melibe leonina} (depicted during swimming in Fig.~\ref{fig0a}(a)) and {\em Dendronotus iris} \cite{NSLGK012,SNLK11,SK11a}.  Our greater goal is to create dynamical foundations for the onset, morphogenesis and structural robustness of rhythmic activity patterns produced by swim CPGs in these animals.  A pilot mathematical model of the {\it Melibe} swim CPG will be discussed in this paper. The circuitry shown in Fig.~\ref{fig0b}(a) depicts only some core elements identified in the biological CPG; its detailed diagram can be found in \cite{sck2014}.   

\begin{figure*}[!t]
  \centering
  {\resizebox*{.7\columnwidth }{!}{\includegraphics{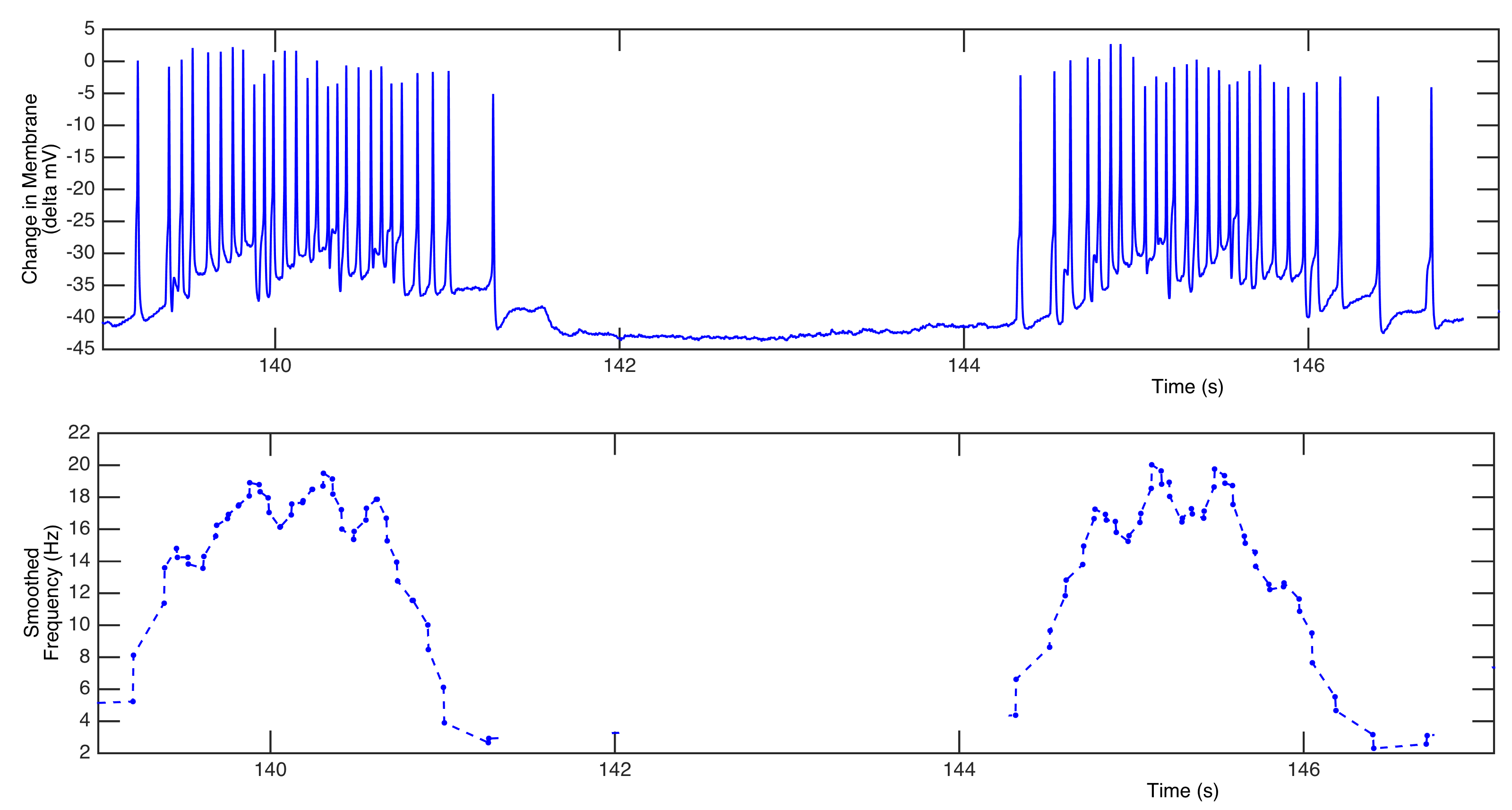}}} 
      \caption{(a) Parabolic distribution of spike frequency within bursts produced by networked interneurons in the {\it Melibe} swim CPG. Recording provided courtesy of A.~Sakurai and time series analysis by  A. Kelley.} \label{fig1}
\end{figure*}

Being inspired by experimental studies of voltage activity recorded from the swim CPGs of the sea slugs {\it Melibe leonina} and {\it Dendronotus iris}, we would like to develop an assembly line for CPG constructors made of coupled biophysically plausible models. Our first simplifying assumption is that CPGs are made of universal building blocks -- half center oscillations (HCOs) \cite{Hill-VanHooser-Calabrese-03}. Loosely speaking, a HCO is treated as a pair of interneurons interacting with each other through reciprocally inhibitory synapses and exhibiting anti-phase bursting.  The interneurons of a HCO can be endogenous bursters, tonic spiking or quiescent ones, which exhibit alternating bursting only when they inhibit each other. Theoretical studies~\cite{Wang-Rinzel-92} have indicated that formation of an anti-phase bursting rhythm is always based on slow subsystem dynamics.  There are three basic mechanisms to generate alternating bursting in the HCO: release, escape, and post-inhibitory rebound (PIR). The first mechanism is typical for endogenously bursting neurons \cite{Jalil-Belykh-Shilnikov-10,Jalil-Belykh-Shilnikov-12}. The other two mechanisms underlie network bursting in HCOs comprised of neurons, which are  hyperpolarized quiescent in isolation \cite{Perkel-Mulloney-74,Skinner-Kopell-Marder-94, Angstadt-Grassmann-Theriault-Levasseur-05,Kopell-Ermentrout-02}. Our second assumption is that the swim CPG interneurons are intrinsic tonic spikers that become network bursters only when externally driven or coupled by inhibitory synapses, as recent experimental studies suggest \cite{sck2014}. The third assumption is that network bursting in the {\it Melibe} swim CPG is {\it parabolic}, i.e. the spike frequency within a burst increases at the middle, and decreases at the ends, as one can observe from  Fig.~\ref{fig1}. This observation indicates the type of neuronal models to be employed to describe network cores. Our model of choice for parabolic bursting is the Plant model \cite{plant75,plant76,plant81}. The Plant model has been developed to accurately describe the voltage dynamics of the R15 neuron in a mollusk {\it Aplysia Californica}, which has turned out to be an endogenous burster  \cite{LL88}.  Most dynamical properties of the R15 neuron have been modeled and studied in detail \cite{CCB91,bertran1993,BCCBB95,butera98,R15,plant_jzlz2015}.

\section{Methods: the Plant model of parabolic bursting} \label{sec:2}

\begin{figure}[!th]
	\centering
\subfigure [][] {\resizebox*{.325\columnwidth}{!}{\includegraphics{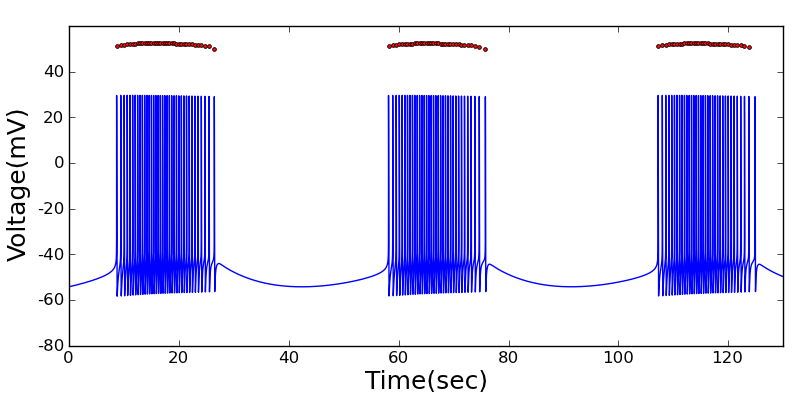}}} 
\subfigure [][] {\resizebox*{.325\columnwidth}{!}{\includegraphics{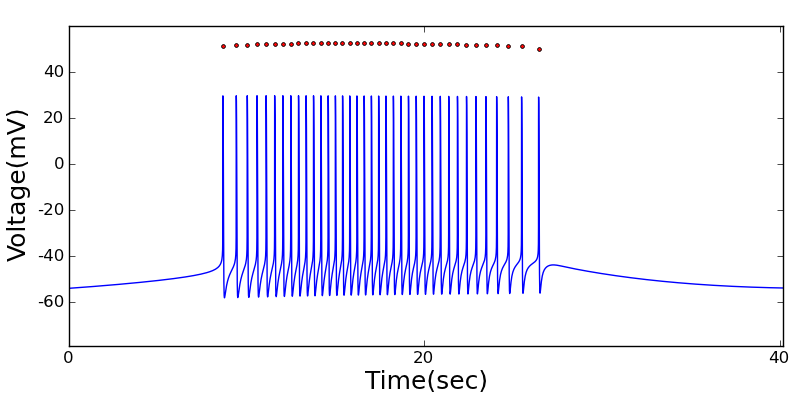}}} 
\subfigure [][] {\resizebox*{.325\columnwidth}{!}{\includegraphics{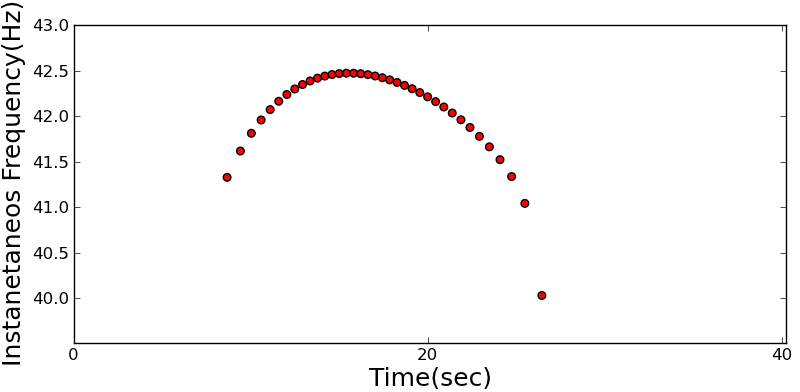}}} 
\caption{(a) Endogenous bursting in the Plant model as alternations of tonic spiking activity and quiescent periods. (b) Single burst featuring a characteristic spike frequency increase in the middle of each burst. (c) Parabolic shape of the  frequency distribution of spikes within a burst is a feature of this kind of bursting. The parameters are $\rho = 0.00015ms^{-1}$, $K_{c}=0.00425ms^{-1}$ and $\tau_{x}=9400ms$.}
	\label{fig2}
\end{figure}

The conductance based Plant model \cite{plant81} for the R15 neuron \cite{R15} located in the abdominal ganglion of a slug {\em Aplysia Californica} is given by the following set of ordinary differential equations derived within the framework of the Hodgkin-Huxley formalism to describe the dynamics of the fast inward sodium [Na], outward potassium [K], slow TTX-resistant calcium [Ca] and an outward calcium sensitive potassium [KCa] currents: 
\begin{equation}\label{eq}
 	C_{m} \dot{V} = -I_{Na} - I_{Ca} - I_{KCa} -  I_{leak} - I_{ext} -I_{syn}. ~\\
 \end{equation}
The last three currents are the generic ohmic leak $I_{leak}$, external constant  $I_{ext}$ and synaptic $I_{syn}$ currents flowing from a pre-synaptic neuron. The full details of the representation of the currents employed in the model are given in the Appendix below. 

\begin{figure*}[b!]
\centering
\includegraphics[width=.46\columnwidth]{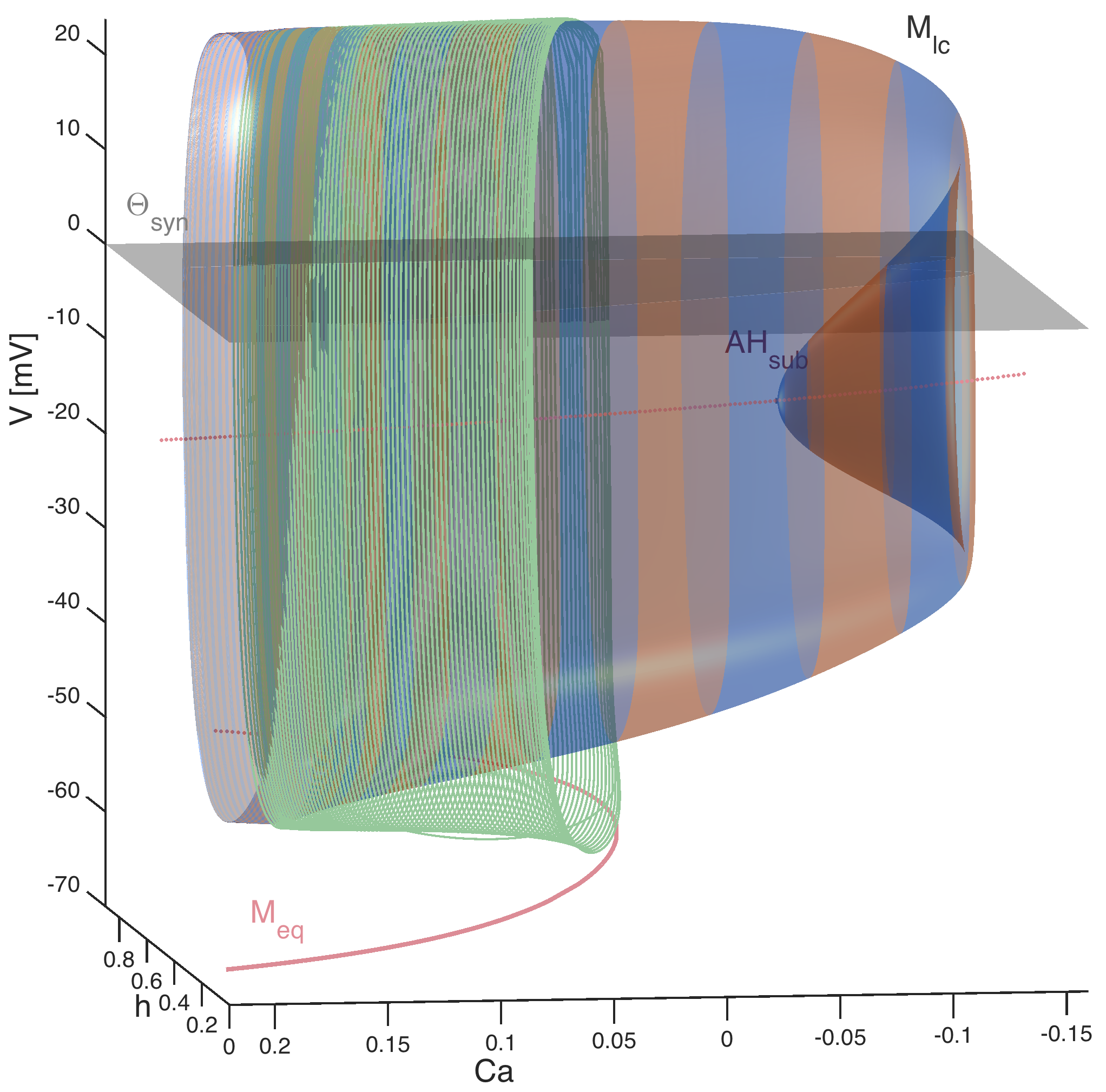}\\
\caption{
Bursting (green) orbit recursively switching between two slow--motion critical manifolds: tonic spiking, $M_{lc}$, with a characteristic fold and originating through a sub-critical Andronov-Hopf (AH) bifurcation from a depolarized equilibrium state,  and quiescent, $M_{eq}$ (orange curve), projected onto the ($h,V$) and slow $Ca$ variables of the of the Plant model; a plane represents the synaptic threshold, $\Theta_{syn}=0mV$.}\label{man}
\end{figure*}

There are two bifurcation parameters in the individual model. The first one is the constant external current, $I_{ext}$, which is set $I_{ext}=0$. Following \cite{Shilnikov-12}, the other bifurcation parameter, $\Delta$, is introduced in the slowest equation:
\begin{equation}
	\quad \dot{Ca} = \rho \left (K_{c} x (V_{Ca}-V+ \Delta)-Ca \right )  \label{sm}
\end{equation}  
describing the concentration of the intracellular calcium in the Plant model. By construction, $\Delta$ is a deviation from a mean value of the reversal potential $V_{Ca}=140mV$ evaluated experimentally for the calcium current in the R15 cells. As such, this makes $\Delta$ a bifurcation parameter. Secondly its variations  are not supposed to alter the topology of the slow motion manifolds in the 5D phase space,  which are called tonic spiking and quiescent in the mathematical neuroscience context, as they are made of, respectively, round periodic orbits and equilibrium states [of the slow subsystem] of the model  (Fig.~\ref{man}).  

At $\Delta=0$, the neuron is an endogenous burster, see Fig.~\ref{fig2}. According to \cite{RL87}, this type of bursting is termed  {\it parabolic.} The reason for this term is that the spike frequency within bursts is maximized in the middle of bursts and minimized at the beginning and the end (see Fig.~\ref{fig2}c). The parabolic structure of a burst is due to the calcium-activated potassium current. Its magnitude is determined by  the intracellular calcium concentration. As the intracellular calcium concentration increases, the calcium dependent potassium current gets activated, which causes an increase of the inward potassium current. As the membrane potential increases over a threshold value, the intracellular calcium concentration decreases, as well as the inward potassium current (see Eq.~(20) in the Appendix). The parabolic distribution of spikes within bursts is shown in Fig.~\ref{fig2}.  The instant frequency value is calculated by the reciprocal of each inter-spike interval. Panels b and c of Fig.~\ref{fig2} clearly disclose the parabolic inter-spike structure of bursts. 

\begin{figure*}[t!]
 \begin{center}
\includegraphics[width=0.99\textwidth]{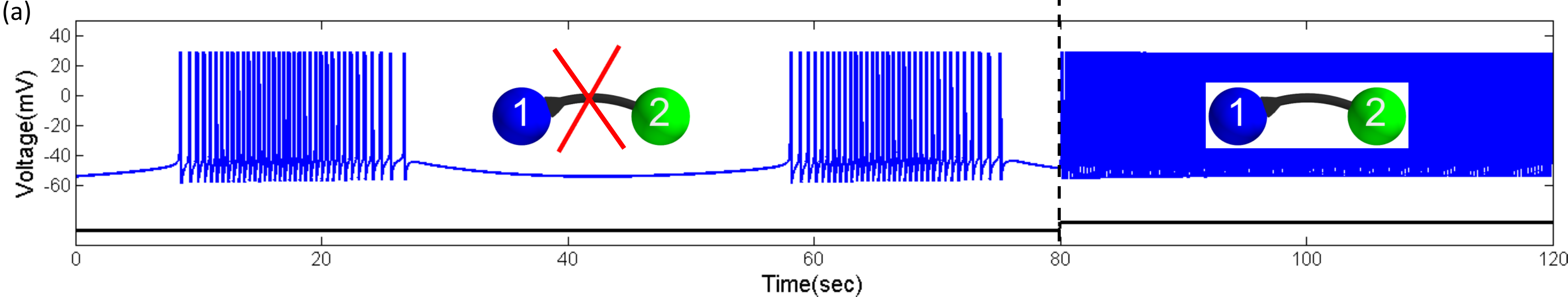}
\includegraphics[width=0.99\textwidth]{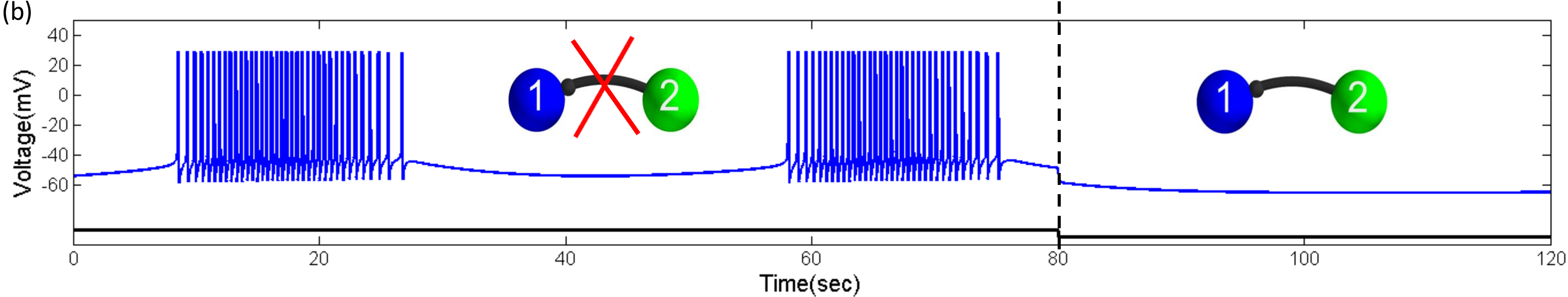}
\caption{Responses of the bursting neuron ($\Delta=0mV$) on the synaptic drive $I_{syn}=g_{syn}(V-V_{rev})$. (a) Excitatory synaptic drive with $g_{syn}=0.002nS$ and $V_{rev}=40mV$ applied at $t=80$sec switches the neuron from bursting to tonic spiking activity. (b) The inhibitory drive with $g_{syn}=0.005nS$ and $V_{syn}=-80mV$ halts bursting and makes the neuron hyperpolarized quiescent.}\label{fig4}
 \end{center}
\end{figure*}

It was shown in \cite{RL87} that the mechanism underlying a transition between quiescent and tonic spiking of bursting in the Plant model is due to a homoclinic bifurcation of a saddle-node equilibrium state  \cite{LP63,LP2014}. This bifurcation occurs in the fast 3D $(V,h,n)$-subspace of the model and is modulated by the 2D slow dynamics in the $(Ca,x)$-variables, which are determined by slow oscillations of the intracellular calcium concentration \cite{plant75,plant76}. The unfolding of this codimension-one bifurcation includes an onset of a stable equilibrium, which is associated with a hyperpolarized phase of bursting, and on the other end, an emergent stable periodic orbit that is associated with tonic spiking phase of bursting. The period of this stable orbit decreases, as it moves further away from the saddle-node  equilibrium mediated by decreasing calcium concentration. The period of the tonic spiking orbit grows with no upper bound as it approaches the homoclinic loop of the saddle-node \cite{SSTC}.      
 
Variations of $\Delta$ change the duty cycle of bursting, which is a ratio of the active tonic spiking phase of bursting to its period. Decreasing $\Delta$ reduces the inactive, quiescent phase of bursting, i.e. increases its duty cycle. Zero duty cycle is associated with the homoclinic saddle-node bifurcation that makes the neuron hyperpolarized quiescent. This corresponds to an emergence of stable equilibrium state for all dynamical variables of the model~(\ref{eq}). In other words, decreasing $\Delta$ makes the active phase longer, so that below a threshold  $\Delta=-32mV$  the neuron switches to tonic spiking activity. Tonic spiking activity is associated with the emergence of a stable periodic orbit in the fast $(V,h,n)$-subspace, while the $(Ca,x)$-variables of the slow subspace converge to a stable equilibrium state. As such, bursting occurs in the Plant and similar models due to relaxation of periodic oscillations in the 2D $(Ca,x)$-subspace, which slowly modulates fast tonic spiking  oscillations in the  $(V,h,n)$ variables. The relaxation limit cycles emerge from one and collapse into the other equilibrium state in the $(Ca,x)$-plane through Andronov-Hopf bifurcations, which can be sub- or super-critical. At the transitions between bursting and tonic spiking, and bursting and hyperpolarized quiescence, the neuron can produce chaotic dynamics, which are basically due to the membrane potential oscillatory perturbations of  plain canards at the folds of the relaxation cycle.

 \begin{figure*}[t!]
 \begin{center}
\includegraphics[width=0.99\textwidth]{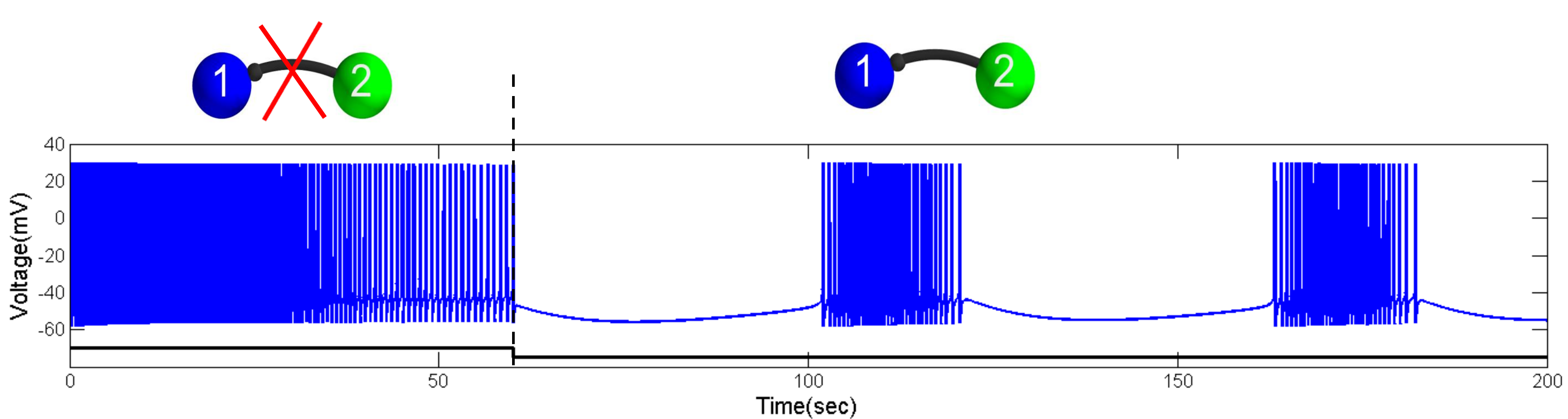}
\caption{Tonic spiking neuron~1 at $\Delta=-34mV$ near the bifurcation transition between tonic spiking and bursting is forced to become a network burster with an application of an inhibitory drive with $g_{syn}^{inh}=0.001nS$, from the pre-synaptic neuron~2  at $t=60sec$. Halting the inhibitory drive restores tonic spiking activity in the targeted neuron (not shown).}\label{fig5} 
 \end{center}
\end{figure*}

\section{Endogenous and network bursting. Inhibitory and excitatory drives}

A half-center oscillator is a network of two neurons coupled by reciprocally inhibitory synapses  that robustly produces 
bursting in alternation, or anti-phase bursting. Such a network can be multistable, i.e. produce other bursting rhythms as well, such as synchronous bursting \cite{Jalil-Belykh-Shilnikov-10} and rhythmic outcomes with slightly shifted phase lags between the endogenously bursting neurons \cite{Jalil-Belykh-Shilnikov-12}. 

In this study, the synaptic current $I_{syn}$ is modeled through the fast threshold modulation (FTM) approach \cite{FTM1}. 
The synapses are assumed to be fast and non-delayed, which is true for the swim CPG in both sea slugs under consideration. 
The synaptic current is given by  
\begin{equation}
	\quad I_{syn} =  g_{syn} (V_{post}-E_{syn}) \frac{1}{1+e^{-k(V_{pre}-\Theta_{syn})}},  \label{syn}
\end{equation}
where $g_{syn}$ is the maximal conductance of the current, which is used as a bifurcation parameter of the networked model; $V_{post}(t)$ and $V_{pre}(t)$ are the voltages on the post-synaptic (driven) and pre-synaptic (driving) neurons; $E_{syn}$ is the synaptic reversal potential. To make $I_{syn}$ excitatory, we set $E_{syn}=40mV$, while in the inhibitory case we set $E_{syn}=-80mV$. 
In Eq.~(\ref{syn}), the second term is a Boltzmann coupling function that quickly, ($k=100$), turns the synaptic current on and off as soon the voltage, $V_{pre}$, of the (driving) pre-synaptic cell(s) raises above and falls below the synaptic threshold, here $\Theta_{syn}=0mV$ (Fig.~\ref{man}).

To model the constant synaptic drive onto the post-synaptic neuron, we assume that $V_{pre}>\Theta_{syn}$. This allows us to calibrate the state of the post-synaptic neuron, and to determine the drive threshold that separates the qualitatively distinct states of the individual and networked neurons. This statement is illustrated in Fig.~\ref{fig4} by simulating responses of the  endogenous parabolic burster to network perturbation. Figure~\ref{fig4}(a) shows, with a properly adjusted excitatory drive, that the endogenous burster switches into tonic spiking activity.  On the other hand, bursting in the networked neuron can be halted when it receives a sufficient inhibitory drive from the pre-synaptic neuron of the network (Figure~\ref{fig4}(b)). Eliminating either drive makes the post-synaptic neuron return to its natural state, i.e. these experiments {\it de-facto} prove that the neuron is mono-stable for the given parameter values. 
 
 A HCO, in the canonical Brown definition \cite{brown}, is a pair of neurons bursting in anti-phase when they are networked by inhibitory synapses. In isolation, such neurons are not endogenous bursters but tonic spikers instead, or remain quiescent \cite{Marder-Calabrese-96}. There are multiple mechanisms underlying such anti-phase bursting, or, more accurately,  anti-phase oscillations  in HCOs and CPGs made of relaxation oscillators \cite{Kopell-Ermentrout-02,Daun-Rubin-Rybak-09}. The list includes the well studied mechanisms of post-inhibitory rebound and escape for quiescent neurons  \cite{Perkel-Mulloney-74,Wang-Rinzel-92, Skinner-Kopell-Marder-94,DESTEXHE-CONTRERAS-SEJNOWSKI-STERIADE-94,Matveev-Bose-Nadim-Farzan-07}, as well as less-known mechanisms of HCOs constituted by intrinsically spiking neurons. Such networks utilizing the Plant models are discussed below.     

To construct such a HCO with relatively weak inhibitory coupling,  the Plant model must be first set into the tonic spiking mode. This is done by setting the bifurcation parameter, $\Delta=-34mV$, see Fig.~\ref{fig5}. Next, we consider a unidirectional network where the tonic spiking neuron~1 starts receiving, an inhibitory drive of $g_{syn}=0.001nS$ from the post-synaptic neuron~2 at $t=60sec$. The inhibitory drive is sufficient to shift the post-inhibitory neuron over the bifurcation transition back into bursting activity. The minimal inhibitory drive must be increased proportionally to make the targeted neuron a network burster whenever it stays further away from the bifurcation transition between tonic spiking and bursting in isolation.

\begin{figure}[t!]
 \begin{center}
\includegraphics[width=0.99\textwidth]{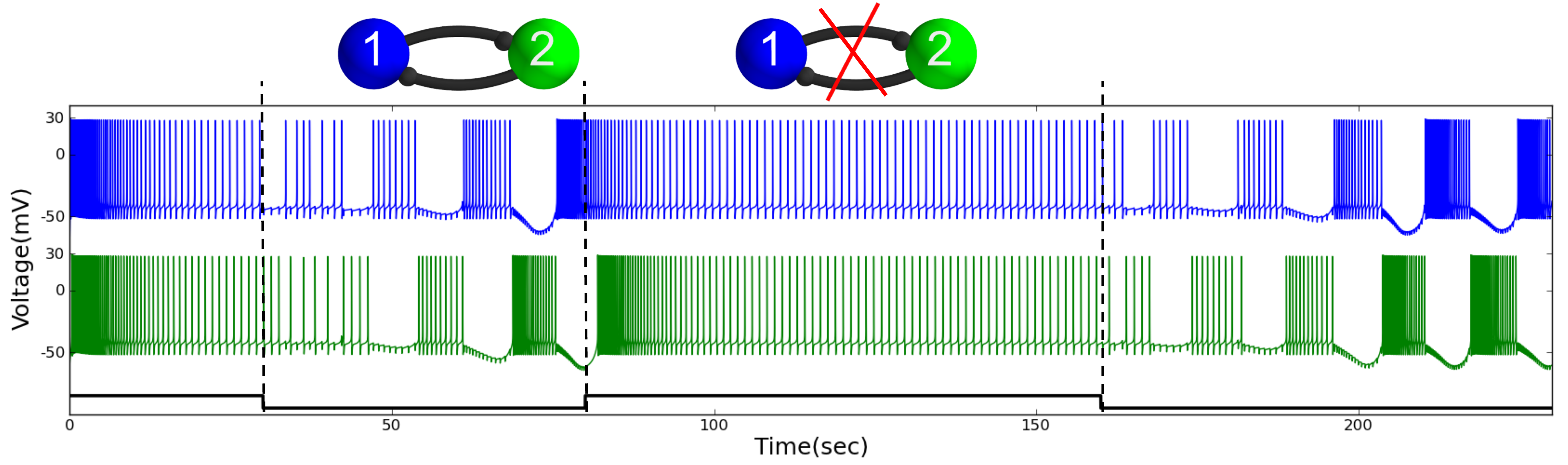}
\caption{Anti-phase network bursting produced by a HCO of two Plant neurons as soon as 
the inhibition is turned on. Blocking the inhibition restores tonic spiking activity in both neurons, and {\it vice versa}. Here, the the network parameters are $g_{syn}^{inh}= 0.008nS$ and  $E_{syn}=-80mV$, and the parameters of the individual neurons are the following: $ \Delta= -60mV, \rho = 0.0003ms^{-1}$, $K_{c}=0.0085ms^{-1}$, $\tau_{x}=235ms$ and $x_{\infty}(V) = 1/(1+e^{-0.15(V+50)})$.
 }\label{fig6}
 \end{center}
\end{figure}

\section{Forming a half-center oscillator}

In this section, we discuss the dynamics of half-center oscillators made of two tonically spiking Plant neurons reciprocally coupled by inhibitory synapses. As before, we describe such synapses within the framework of the fast threshold modulation (FTM) paradigm using Eq.~(\ref{syn}) to match the shape and magnitude  of inhibitory postsynaptic potentials (IPSPs) in the post-synaptic neurons. IPSPs are the indicators of the type and the strength of synapses in the network. 
  
We perform simulations in a fashion that is analogous to the dynamic clamp technique used in neurophysiological experiments. The approach involves the \emph{dynamic} block, restoration and modulation of synaptic connections during simulation. These modeling perturbations should closely resemble the experimental techniques of drug-induced synaptic blockade, modulation, wash-out, etc. Restoring the chemical synapses during a simulation makes the HCO regain network bursting activity with specific phase characteristics. Depending on the coupling strength as well as the way the tonically spiking neurons are clamped, the network bursting may change phase-locked states, i.e. be potentially multi-stable. Experimental observations also suggest specific constraints on the range of coupling strengths of the reciprocal inhibition, such that the networks  stably and generically achieve the desired phase-locking. 
 
\begin{figure*}[t!]
\centering
\includegraphics[width=0.99\textwidth]{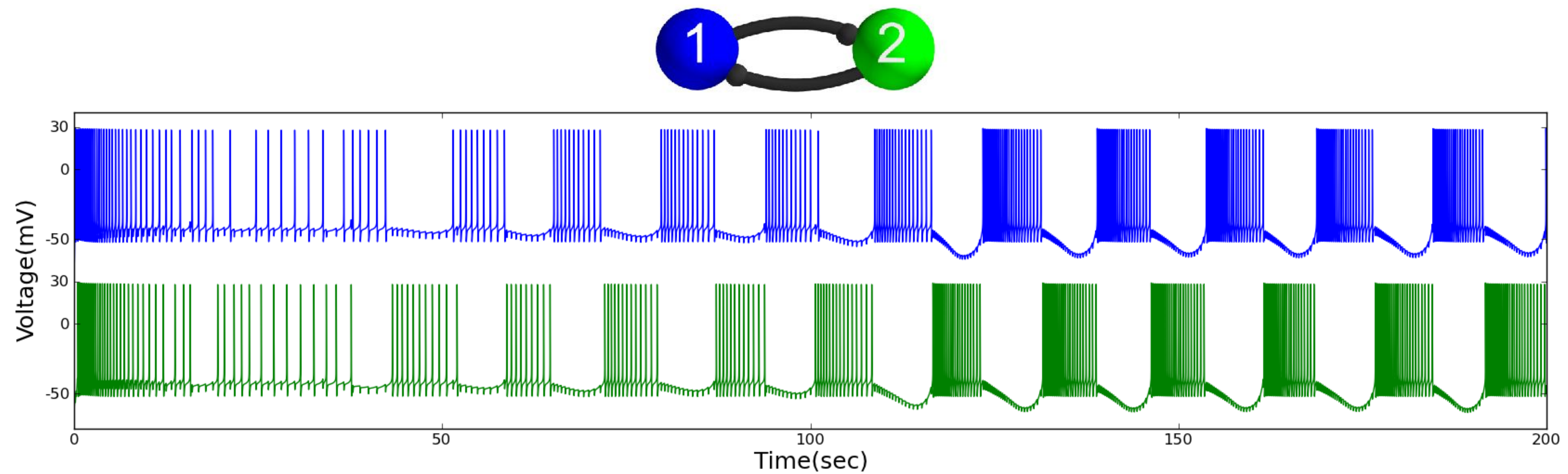}
\caption{Onset of emergent network anti-phase bursting in the HCO with reciprocally inhibitory. 
($E_{syn}=-80mV$) synapses at $g^{inh}_{syn}=0.0073nS$. 
}\label{fig7}
\end{figure*}

Figure~\ref{fig6} demonstrates the stages of anti-phase bursting formation in the HCO. The uncoupled neurons are initiated in tonic spiking mode. After turning on the reciprocally inhibitory synapses $g_{syn}= 0.008nS$, the HCO quickly transitions to the regime of robust anti-phase bursting. Turning off the synapses restores the native tonic spiking activity in both neurons. Turning on the reciprocal synapses makes the HCO regain the network bursting. Note that the length of transients from tonic spiking to network bursting depends on the strength of the synaptic coupling for the fixed parameters of the individual Plant neurons. By comparing the magnitude of IPSPs in the voltage traces represented in Figs.~\ref{fig6} and \ref{fig7}, one can conclude that the coupling in the later case is weaker. This is why, the onset of network bursting in the HCO is less pronounced. 

Our modeling studies agree well with experimental recordings from the identified interneurons in the {\it Melibe} swim CPG which suggests that the observed bursting is due to synergetic interactions of interneurons of the network \cite{sck2014}. One can see from Fig.~\ref{fig0a}(b) that network bursting in the biological HCO formed by  two Si3 interneurons of the {\it Melibe} swim CPG  is seized as soon as the right one, Si3R, receives a negative current pulse that makes it hyperpolarized quiescent, while its left bursting counterpart, Si3L, turns into tonic spiking activity instead. Moreover, one can deduct from the wiring diagram of the CPG depicted in  Fig.~\ref{fig0b}(a) and the analysis of voltage traces represented in Fig.~\ref{fig0a}(b) that the interneuron Si2L becomes a tonic-spiker as soon as the pre-synaptic interneuron Si3R stops inhibiting it (compare with Fig.~\ref{fig6}.)  This further supports the assertion that the swim CPG is made of intrinsically tonic spiking interneurons.

\begin{figure*}[b!]
\centering
\includegraphics[width=0.99\textwidth]{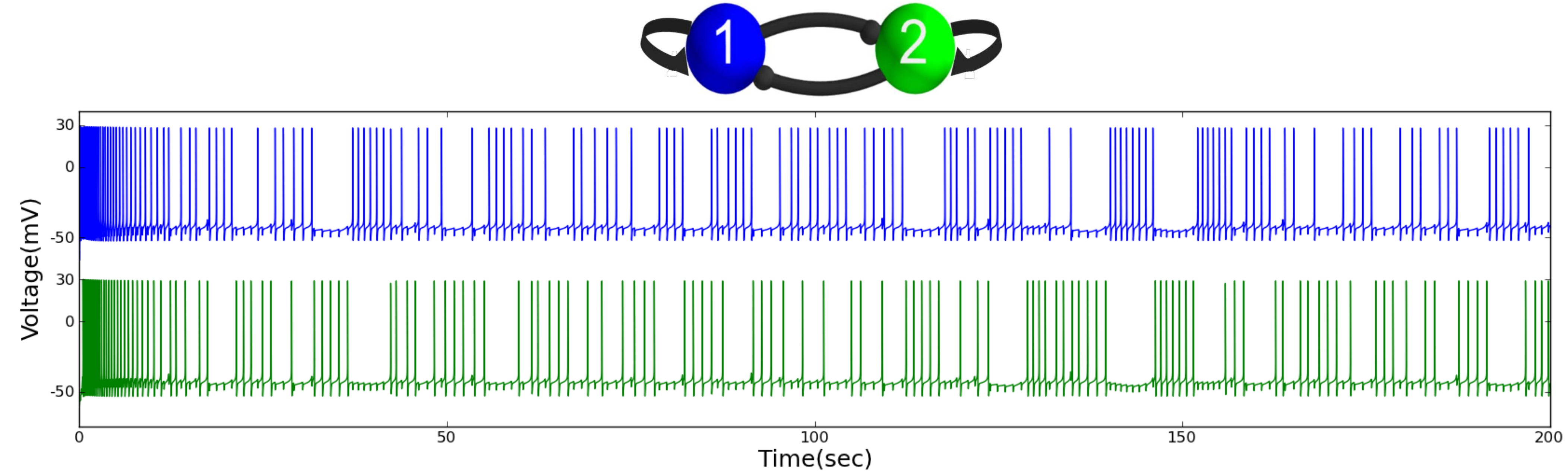}
\caption{Turning on the excitatory autapses at $g_{aut}^{exc}=0.016nS$ in the HCO with $g_{syn}^{inh}=0.0073nS$ halts pronounced network bursting.}
\label{fig8}
\end{figure*}


To test the robustness of network anti-phase bursting to perturbation and to calibrate the necessary influx of reciprocal inhibition generated by the Plant neurons, we consider a HCO with excitatory autapses. The objective here is to determine an equivalent amount of excitatory drive to be projected onto the post-inhibitory network burster to cancel out the inhibitory drive and  shift it back to the initial tonic-spiking mode. 

An autapse is a synapse of a neuron onto itself, where the axon of the neuron ends on its own dendrite.  After their discovery \cite{autopse} autapses have been observed in a range of  nervous systems. The autapses are arguably to be responsible for tuning of neural networks. 
This particular configuration of the HCO depicted in Fig.~\ref{fig8} is formally motivated by the swim CPG circuitry, see Fig.~\ref{fig0b}(a). One can see from it that the interneurons of the bottom HCO receive excitatory drives from the top interneurons forming the top HCO. We would like to find the threshold over which the neurons no longer form a stably bursting HCO. This would allow us to calibrate and quantify the relative strengths of the mixed synaptic connections in the swim CPG models.     

In this HCO configuration, each neuron inhibits its counterpart and self-excites through the autapse. Both autapses are introduced to the model using the FTM approach with $E_{aut}=40mV$. In this experiment, the conductance values for inhibitory synapses  are set at $g_{syn}^{inh}=0.0073nS$. This is sufficient for the HCO to generate robust anti-phase bursting as seen in Fig.~\ref{fig7}. Next, we add the autapses along with inhibition and gradually increase $g_{aut}^{exc}$. We found that increasing $g_{aut}^{exc}$ proportionally increase the delay. At $g_{aut}^{exc}=0.016nS$, the network stops exhibiting anti-phase bursting. We note that unlike a permanent excitatory drive from pre-synaptic neurons, an introduction of the excitatory autapse, acting only when the self-driving neuron is above the synaptic  threshold, is effectively perturbation equivalent for the calibration purpose.       

\section{Assembly line of a {\it Melibe} swim CPG}

\begin{figure*}[t!]
\centering
\includegraphics[width=0.75\textwidth]{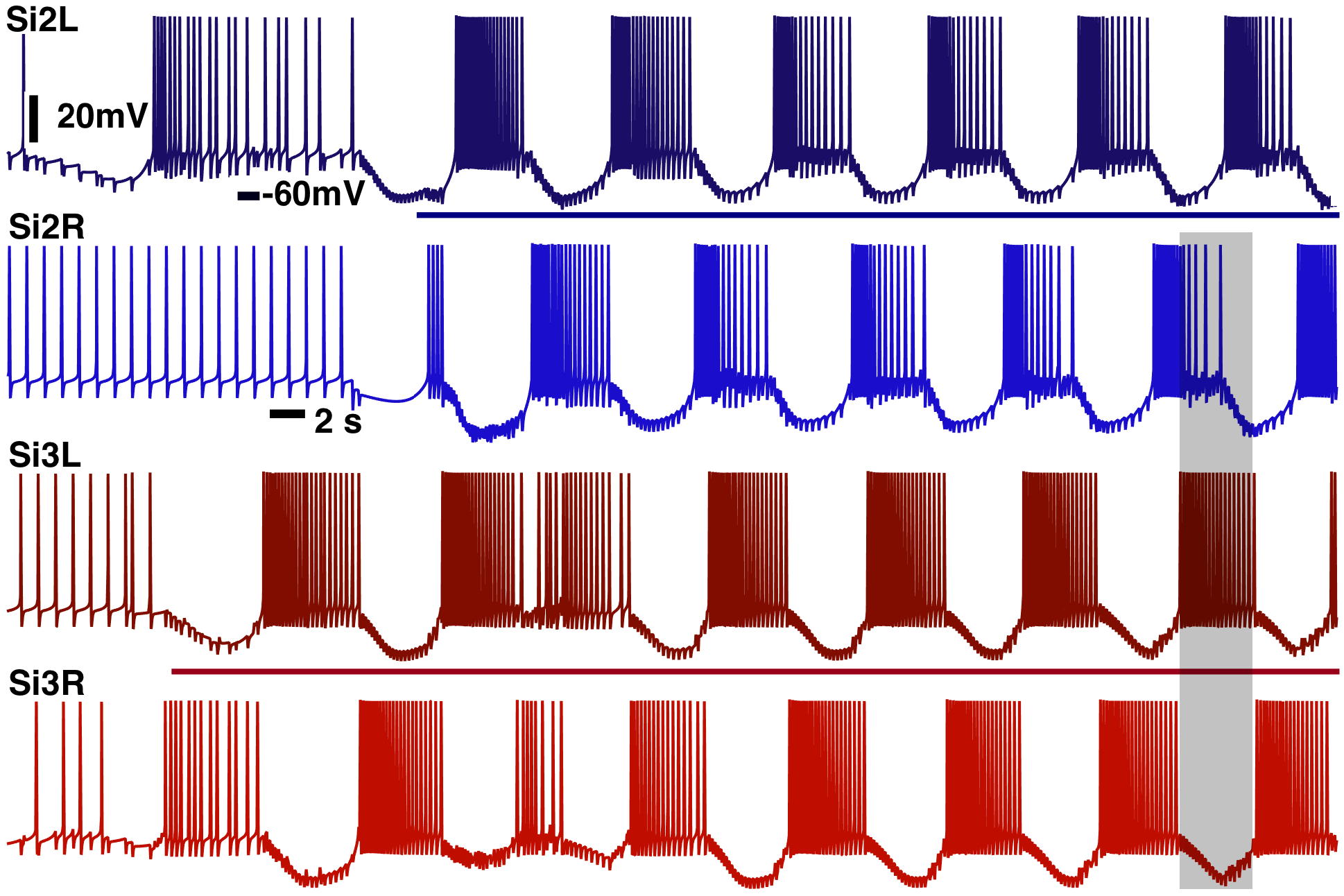}
\caption{Assembly line of the {\it Melibe} swim CPG model out of four intrinsically tonic spiking Plant neurons. First the reciprocal inhibition between Si3R and Si3L is turned on, followed by turning on the reciprocal inhibition between  Si2R and Si2L, and next simultaneous turning on unidirectional cross-lateral inhibition from Si3R(L) projected onto Si2L(R), and  bi-lateral excitation originating from Si2R(L) down onto Si3R(L).  After a short transient, the CPG model exhibits the desired 3/4 phase shift lag between Si2L and Si3L.  Compare with voltage traces of the biological CPG in Fig.~\ref{fig0b}(b).}\label{fig9}
\end{figure*}

In this final section, we put together a pilot model of the {\rm Melibe} swim CPG according to a circuitry based on identified interneurons and synapses; its wiring diagram is sketched  in Fig.~\ref{fig0b}(a). This network model is made of the two HCOs constituted by tonic spiking Plant neurons. We would like to find out whether this sample CPG model can already produce  phase lags similar to those between bursting interneurons in the biological CPG. For the sake of simplicity, we do not include Si4R/L interneurons in the model and we also omit electrical synapses. It is known from experimental studies  \cite{sck2014} that blocking chemical, inhibitory and excitatory synapses between the interneurons may be sufficient to break down the motor pattern by the network.   
Figure~\ref{fig0b}(b) points out that the interneurons of either HCO burst in anti-phase and there is the characteristic 3/4 phase lag between the burst initiation in the neurons Si2L and Si3L, as well as between Si2R and Si3L. This phase lag is repeatedly observed in both adult and juvenile animals.

As before, we use the Plant neurons initiated in the tonic spiking mode, relatively close to the transition to bursting. Initial conditions of the neurons are randomized. After letting the neurons settled down to tonic spiking activity, the network connections are turned on. As Fig.~\ref{fig9} shows, with the reciprocal inhibition being first turned on, the bottom interneurons Si3L and Si3R become anti-phase network bursters, and so do  Si2R and Si2L as soon as the reciprocal inhibition between them is turned them on, too. At this stage, the CPG model is formed by two uncoupled HCOs. 
 A few seconds later, they become coupled by simultaneous turning on the unidirectional cross-lateral inhibition from Si3R(L) projected onto Si2L(R), and  bi-lateral excitation from Si2R(L) down onto Si3R(L). One can see from this figure
 that all four interneurons of the  CPG model exhibit network bursting with the desired phase lags. These are 0.5 (half period) between the interneurons of each HCO, and 3/4 (a fraction of the network period) between the HCOs, or between the corresponding reference interneurons Si2L and Si3L. We note that such a phase shift was reported in a similar {\it Melibe} swim CPG constituted by endogenous bursters; that model also incorporated electrical synapses \cite{jalil2013}. There is a great room for improvement of CPG network models to include other identified interneurons and to incorporate additional electrical synapses to find out whether additions of new elements can stabilize or desynchronize the desired bursting pattern as it was done using the Poincar\'e return maps for endogenous bursters \cite{Wojcik-Schwabedal-Clewley-Shilnikov-14}.   Of our special interest are various problems concerning structural stability of the network, and its robustness (Lyapunov stability) for bursting outcomes subjected to perturbations by pulses of the external current, as well as reductions to return maps between burst initiations in constituent neurons.  These questions are beyond the scopes of the given examination and will be addressed in full detail in our forthcoming publications soon. The question about a possible linking of the characteristic $\frac{3}{4}$ phase lag and the {\it Melibe leonina} lateral swim style is the paramount one among them. 
   
\section{Summary}

We have discussed a basic procedure for building network bursting CPGs made of intrinsically tonic spiking neurons. As a model for such networks, we have employed the biophysically plausible Plant model that was originally proposed to describe endogenous bursting R15-cells in the {\it Aplysia} mollusk. Such bursting was  intracellularly recorded,  and identified as parabolic,  from the known interneurons in the swim CPGs of two sea slugs: {\it Melibe leonina} and {\it Dendronotus iris}. There is experimental evidence that bursting in these swim CPGs is due to synergetic interactions of all constituent neurons that are intrinsic tonic-spikers in isolation. To model the {\it Melibe} swim CPG, we have first examined dynamical and structural properties of the Plant model and its responses to perturbations. These perturbations include inhibitory and excitatory inputs from pre-synaptic neurons in the network.  We have identified  the transition boundary beyond which the bursting Plant model becomes a tonic-spiker and shifted it slightly over the threshold using an introduced bifurcation parameter. We have shown that the perturbed/calibrated Plant neuron, exhibiting intrinsically tonic spiking activity, becomes a network burster when it receives an inhibitory drive from a pre-synaptic neuron. By combining two such neurons, we have created a genuine half-center oscillator robustly producing anti-phase bursting dynamics. We have also considered a HCO configuration with two excitatory autapses to assess the robustness of anti-phase bursting with respect to excitatory perturbations. Finally, we have employed all necessary components to assemble a truncated model of the {\it Melibe} swim CPG with the characteristic 3/4-phase lags between the bursting onsets in the four constituent interneurons. In future studies, we plan to examine the dynamics of the CPG models with all synaptic connections, including electrical, as well as incorporating additional identified interneurons.  We will also explore their structural stability, robustness and potential multi-stability of their bursting outcomes with various phase lags. An additional goal is to find out whether the motor pattern  with the 3/4-phase lags will persist in networks 
with interneurons represented by other mathematical models including phenomenologically reduced ones. 
Potentially, these findings shall provide a systematic basis for comprehension of plausible biophysical mechanisms for the origination and regulation of rhythmic patterns generated by various CPGs.  Our goal is to extend and generalize the  dynamical principles disclosed in the considered networks for other neural systems besides locomotion, such as olfactory cellular networks.  

\section{Acknowledgments}

A.S. also acknowledges the support from GSU Brain and Behaviors pilot grant, RFFI 11-01-00001, RSF grant 14-41-00044 at the Lobachevsky University of Nizhny Novgorod and the grant in the agreement of August 27, 2013 N 02.B.49.21.0003 between the Ministry of education and science of the Russian Federation and Lobachevsky State University of Nizhni Novgorod (Sections 2-4), as well as NSF BIO-DMS grant XXXXX. We thank A. Sakurai and P.S. Katz for sharing experimental data and suggestions, and the acting and associated members of the NEURDS (Neuro Dynamical Systems) lab at GSU: A. Kelley, K. Pusuluri, S. Pusuluri, T.Xing, M. Fen, F. Fen, J. Collens, D. Knapper, A. Noriega, P. Ozluk, B. Chung and E. Latash for helpful discussions, and R.~Clewley for his guidance on the PyDSTool package \cite{PyDSTool} used in simulations.

\newpage

\section{Appendix: the conductance based Plant model}

The model in this study is adopted from \cite{plant81}. The dynamics of the membrane potential, $V$, is governed by
the following equation:
\begin{equation}
 	C_{m} \dot{V} = -I_{Na} - I_K - I_{Ca} - I_{KCa} -  I_{leak} -I_{syn}, \label{a1}
 \end{equation}
where  $C_{m}=1\mu F/cm^2$  is the membrane capacitance, $I_{Na}$ is the $Na^+$ current, $I_{K}$ is the $K^+$ current, $I_{Ca}$ is the $Ca^{+2}$ current, $I_{KCa}$ is the $Ca^{2+}$ activated $K^+$ current, $I_{leak}$ is the leak current, $I_{syn}$ is the synaptic current. The fast inward sodium current is given by   
\begin{equation}
I_{Na}=g_{Na} m^{3}_{\infty}(V)h(V-V_{Na}), 
 \end{equation}
where the reversal potential $V_{Na}=30mV$ and the maximum $Na^+$ conductance value $g_{Na}=4nS$. The instantaneous activation variable  is defined as 
\begin{equation}
m_{\infty}(V)= \frac{\alpha_{m}(V)}{\alpha_{m}(V) +\beta_{m}(V)}, 
\end{equation}
 where  
 \begin{equation}
\alpha_{m}(V) = 0.1 \frac{50-V_s}{\exp((50-V_s)/10)-1} , \quad \beta_{m}(V) = 4 \exp((25-V_s)/18),  
\end{equation}
while the dynamics of  inactivation  variable $h$ is given by      
  \begin{equation}
 \dot{h} =  \frac{h_{\infty}(V)-h}{ {\tau_{h}(V)} },
  \end{equation}
where
\begin{equation}
h_{\infty}(V)= \frac{\alpha_{h}(V)}{\alpha_{h}(V) +\beta_{h}(V)}\quad \mbox{and} \quad \tau_{h}(V) = \frac{12.5}{\alpha_{h}(V) +\beta_{h}(V)}, 
\end{equation}
with  \begin{equation}
 \alpha_{h}(V) = 0.07 \exp((25-V_s)/20) \quad \mbox{and} \quad  \beta_{h}(V) = \frac{1}{\exp((55-V_s)/10)+1},   
\end{equation}  
where
\begin{equation}
 V_s = \frac{127V+8265}{105}mV.
\end{equation}  
The fast potassium current is given by the equation  
\begin{equation}
I_K=g_{K} n^{4}(V-V_{K}),  
\end{equation}
where the reversal potential is $V_{K}=-75mV$ and the maximum $K^+$ conductance value is $g_K=0.3nS$.The dynamics of inactivation gating variable is described by 
\begin{equation}
\dot{n} = \frac{n_{\infty}(V)-n}{\tau_{n}(V)}, 
 \end{equation}
where 
\begin{equation}
n_{\infty}(V)= \frac{\alpha_{h}(V)}{\alpha_{h}(V) +\beta_{h}(V)} \quad \mbox{and} \quad \tau_{n}(V) = \frac{12.5}{\alpha_{h}(V) +\beta_{h}(V)}, 
\end{equation}
with 
\begin{equation}
	\alpha_{n}(V) = 0.01 \frac{55-V_s}{\exp((55-V_s)/10)-1} \quad \mbox{and} \quad  \beta_{n}(V) = 0.125 \exp((45-V_s)/80).
\end{equation}	
The TTX-resistant calcium current is given by  
 \begin{equation}
 I_{Ca} =g_{Ca} x (V-V_{Ca}),  
 \end{equation}
 where the reversal potential is $V_{Ca}=140mV$ and the maximum $Ca^{2+}$ conductance is $g_{Ca}=0.03nS$. The dynamics of the slow activation variable is described by  
 \begin{equation}
\dot{x} = \frac{x_{\infty}(V)-x}{\tau_{x}(V)}, 
\end{equation}	
where 
\begin{equation}
x_{\infty}(V) = \frac{1}{\exp(-0.3(V+40))+1} \quad \mbox{and} \quad \tau_{x}(V)=9400ms.
\end{equation}
The outward $Ca^{2+}$ activated $K^+$ current is given by  
\begin{equation}
I_{KCa} =g_{KCa}\frac{[Ca]_i}{0.5+[Ca]_i}(V-V_{K}),  
\end{equation}
where the reversal potential is $V_{Ca}=140mV$. The dynamics of intracellular calcium concentration is governed by 
\begin{equation}
\dot{Ca}  =  \rho \left [ K_{c} x (V_{Ca}-V)-[Ca]_i \right  ],  
\end{equation}
 where the reversal potential is $V_{Ca}=140mV$, and the constant values are  $\rho= 0.00015mV^{-1}$ and  $K_c=0.00425mV^{-1}$.
The leak current is given by 
\begin{equation}
I_{leak}=g_{L} (V-V_{L}), 
\end{equation}
where the reversal potential $V_{L}=-40mV$ and the maximum conductance value $g_{L}=0.0003nS$.  The synaptic current is defined as 
\begin{equation}
\quad I_{syn} =  \frac{g_{syn} (V_{post}-E_{rev})}{1+e^{-k(V_{pre}-\Theta_{syn})}}  
\end{equation}
with the synaptic reversal potential $V_{post}=-80mV$ for inhibitory synapses and $V_{post}=40mV$ for excitatory synapses and the synaptic threshold $\Theta_{syn}=0mV$, and $k=100$.


\end{document}